\documentclass[pra, 12 pt]{revtex4}
\usepackage[english]{babel}
\usepackage{amsmath}
\usepackage{epsfig}

\begin{document}


\title{\bf UNIVERSALITY OF ROTONS IN LIQUIDS, GENERALIZED SUPERFLUIDITY CRITERION AND HELONS IN HELIUM II
}

\author{V.B.Bobrov $^{1,2}$, S.A.Trigger$^{1,3}$}
\address{$^1$Joint\, Institute\, for\, High\, Temperatures, Russian\, Academy\,
of\, Sciences, 13/19, Izhorskaia Str., Moscow\, 125412, Russia;\\
$^2$National Research University "MPEI"\,,
Krasnokazarmennaya str. 14, Moscow, 111250, Russia;\\
$^3$ Eindhoven  University of Technology, P.O. Box 513, MB 5600
Eindhoven, The Netherlands;\\
emails:\, vic5907@mail.ru,\;satron@mail.ru}

\begin{abstract}
An analysis of experimental data shows that, in addition to phonon--roton excitations in superfluid helium, there necessarily exist at least one branch of elementary excitations whose energy spectrum strongly depends on temperature. On this basis, the Landau superfluidity criterion is generalized for several branches of elementary excitations, taking into account that the critical velocity should vanish during the phase transition of liquid helium from the superfluid state to the normal state. We suppose that a new spectrum of excitations with a gap, depending on interparticle interaction, corresponds to the transition of helium to superfluid state. This gap exists only in superfluid state and disappears at the transition temperature $T_\lambda$. The phonon-roton branch of excitations has no crucial influence on superfluidity. Rotons, as well as phonons, are not the specific excitations for the superfluid helium, but exist as universal excitations in liquid state, and form the continuous branch of excitations in different liquids. This point of view is confirmed experimentally and numerically.

PACS:

05.30.Jp - Boson systems;\,
03.75.Kk - Dynamic properties of condensates; collective and hydrodynamic excitations, superfluid flow;\,
03.75.Nt - Other Bose-Einstein condensation phenomena;\,
05.70.Fh - Phase transitions: general studies;\,
61.20.Gy - Theory and models of liquid structure;\,

Key words: superfluid helium, elementary excitations, phonon-roton excitations, Landau superfluidity criterion

\end{abstract}

\maketitle

\section{Introduction }

More than 70 years have passed since Landau [1, 2] has formulated the phenomenological superfluidity theory. However, the problem of constructing the consistent superfluidity theory satisfying all known experimental facts has not yet been solved (see, e.g., [3--5] and references therein). The main and unchanged postulate of the Landau theory is the statement that superfluid helium is a quantum fluid consisting of a superfluid component moving without friction (and not involved in the energy transfer in the form of heat) and a normal component moving with friction (and involved in heat transfer). Two-liquid hydrodynamics developed on this basis made it possible to explain many experimental data (see, e.g., [6]).

Another no less important postulate of the phenomenological Landau theory [1] is the statement that the normal component in superfluid helium is a gas of elementary excitations characterized by the dependence of the energy spectrum $\varepsilon(p)$ on the momentum $p$. From this statement, it immediately follows that, if the superfluid component velocity reaches the critical velocity $V_{cr}$ defined from the condition
\begin{eqnarray}
V_{cr}=min \left(\varepsilon(p)/p\right), \label{A1}
\end{eqnarray}
superfluidity breakdown occurs. Thus, at velocities $V > V_{cr}$, the superfluidity phenomenon cannot be observed. This statement [1] is known as the Landau superfluidity criterion. In [2], Landau proposed the phonon--roton spectrum $\varepsilon(p)$ of elementary excitations for superfluid helium, whose linear region at small momenta $p$ is attributed to phonons; the maximum and minimum of this curve are attributed to maxons and rotons, respectively.\\
Taking into account the statement by Feynman [7] about the relation of the spectrum of elementary excitations to peak positions of the dynamic structure factor $S(p,\omega)$, the shape 
-----------------\\
This text is the extended version of the paper accepted for publication in Russian journal  "Kratkie soobschenija po fizike" (P.N. Lebedev Physical Institute of Russian academy of science), N 6 (2013). Translated as:  Bulletin of the Lebedev Physics Institiute.\\

of the phonon--roton spectrum of elementary excitations, predicted by Landau, was confirmed in the experiments on inelastic neutron scattering in superfluid helium [8, 9]. \\
The typical parameters of the phonon--roton spectrum in superfluid helium are: the speed of sound, which defines the phonon spectral region, $c\approx 250$ m/s, the roton gap $\Delta^{p-r}\approx 8,6 °K$, and the critical velocity related to the roton gap.
In this case, the critical velocity $V_{cr}^{exp}$  observed in superfluid helium can be hundred times lower than the critical velocity related to the roton gap $V^{(p-r)}_{cr}\approx 60$ m/s (see [10] for more details).
We note that maxima $V_{cr}^{exp}\approx (2\div 3)$ m/s in ultrathin films and capillaries were observed at temperatures $T< 1 °K$ (see [11] for more details); in the case of superfluid helium flowing through holes a few micrometers in diameter in thin partitions, critical velocities $V_{cr}\approx (8 \div1 0)$ m/s were fixed [12]. However, such radical difference of these values from $V_{cr}$ related to roton gap necessitated the search for the other superfluidity breakdown mechanism unrelated to roton excitation.

After the key study by Onsager [13], Feynman [14] proposed the assumption that the quantum of action should enter to the phenomenological Landau theory through the conditions of superfluid component velocity quantization. The corresponding conditions are postulated and were confirmed experimentally [15, 16]. As a result, the superfluidity breakdown during superfluid helium motion is currently associated with the generation of extensive Onsager--Feynman quantum vortices or closed vortex filaments (loops, rings) [11], which, in particular, allows the description of the experimental dependence of the critical velocity on the capillary hole size at which the superfluidity breakdown occurs. However, as noted in [3], such an explanation is inapplicable to the case of ultrathin capillaries in which quantum vortices with the superfluid velocity slowly decreasing with the distance from the vortex axis cannot be generated. Such situation [3] is similar to the problem of critical currents in superconductors of the second type, where Abrikosov quantum vortices do not fit into thin superconducting filaments whose thickness is smaller than the London penetration depth of the magnetic field [17].

In our opinion, to explain the experimental data on the critical velocity, we should take into account the possible existence of at least one more branch of elementary excitations in superfluid helium. As will be shown below, it is not the only cause necessitating the consideration of other branches of elementary excitations.

\section{Branches of elementary excitations}

According to the above discussion, let us generalize the Landau superfluidity criterion formulation, since several branches of elementary excitations can exist in superfluid helium. Then, among all possible critical velocities $V^{(\alpha)}_{cr}$ determined by the shape of the spectrum for each elementary excitation branches $\varepsilon^{(\alpha)}$ (superscript $\alpha$), our interest will be in only those providing minimal value of $V^{(\alpha)}_{cr}$. Hence, $V^{(\alpha)}_{cr}$ entering relation (1) is determined from the condition
\begin{eqnarray}
V_{cr}=min_{(\alpha)} V^{(\alpha)}_{cr}, \qquad V^{(\alpha)}_{cr}=min \left(\varepsilon^{(\alpha)}(p)/p\right), \label{A2}
\end{eqnarray}.
Therefore, we note that Landau in his first paper [1] assumed the existence of two types of elementary excitations in superfluid helium: phonons associated with the potential fluid motion and rotons associated with vortical fluid motion (see Fig. 1 and detailed description in [3--5]). However, within this model, he failed to quantitatively describe the experimental data on the speed of the second sound which was measured with high accuracy by Peshkov [18]. Therefore, in his next paper [2] Landau assumed the existence of only one type of excitations with phonon--roton energy spectrum. In this case, the roton spectral region should not be associated with vortical motion [19].

We also note that numerous attempts of the theoretical prediction of the existence of one more branch of elementary excitations in superfluid helium were undertaken. They are based on the hypothesis of the Bose--Einstein condensate existence in superfluid helium. In particular, this leads to the gap in the energy spectrum of degenerate weakly nonideal boson gas at small momenta associated with Bose--Einstein condensate (see [20--23] and references therein). In the alternative version (see [24] for more details and references therein), the model of nonideal gas with boson pair condensation is similar to Cooper pairing of electrons in superconductors [25].
\begin{figure}[h]
\centering\includegraphics[width=6cm]{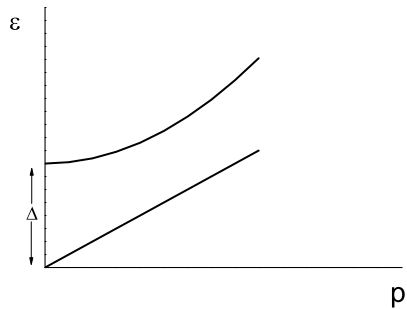}
\caption{{\protect\footnotesize {Spectrum of superfluid $He$ excitations with the gap $\Delta$, proposed in [1]}}}\label{Fig.1}
\end{figure}

At first sight, the development of this theory of superfluid helium as a single-component boson liquid is prevented by the Hugengolz--Pines theorem [26] according to which the gap cannot exist in the energy spectrum of such a system at small momenta. However, it should be kept in mind that the Hugengolz--Pines theorem proof [26] is conceptually based on the  Bogolyubov's statement [27] about the possibility of replacing the boson creation and annihilation operators by C-numbers at zero momentum. In fact, this means that a system with the other Hamiltonian is considered instead of the initial degenerate boson liquid. At the same time, such a theorem has not yet been proved based on the diagram technique by Belyaev for the boson system [28] in which Bogolyubov's statement [27] is not used. In proving the theorem, Hugengolz and Pines [26] "simplified" the diagram technique by Belyaev [28], using the above Bogolyubov's statement [27] (see [29] for more details). We note that the results of [27, 28] can be obtained without the use of "anomalous averages" for degenerate weakly nonideal boson gas [30]. Thus, the validity of the Hugengolz--Pines theorem [26] for the boson system with the initially written Hamiltonian cannot currently be considered as proved.

Furthermore, in the case of the spectral gap, corresponding elementary excitations give an exponentially small contribution to the physical quantities at temperatures close to absolute zero. For the same cause, such excitations are almost absent when considering the zero temperature for which the Hugengolz--Pines theorem was proved [26].

Attention should also be paid to the fact that the Landau superfluidity theory [1, 2] is unrelated to the hypothesis of the Bose--Einstein condensate existence (see [5, 6] for more details). Thus, we have no any constraints for the statements concerning the existence of several types of elementary excitations in superfluid helium.
\begin{figure}[h]
\centering\includegraphics[width=6cm]{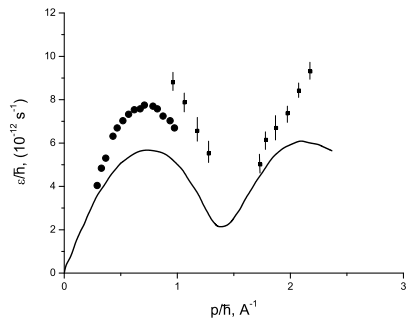}
\caption{{\protect\footnotesize {The position of the maximum in the dynamic ion structure factor in liquid rubidium at 320 K: full circles are the experimental data of Copley and Rowe [43], vertical bars are the experimental data of Glaeser et al. [44], solid curve is the numerical calculation of the dispersion equation [45].}}}\label{Fig.2}
\end{figure}

From this point of view, the papers by Glyde and Griffin [31, 32] based on the experimental data [33--35] should be noted. They put forward the assumption on the different physical nature of the phonon and maxon--roton spectral regions, proposed by Landau [2]. According to Glyde and Griffin, the phonon spectral region characterizes collective sound excitations, while the maxon--roton region is controlled by single-particle excitations. The single curve results from "hybridization" of these two excitation branches (see also [35, 36]).

However, the experiments on inelastic neutron scattering [34, 35, 38] show that the phonon--roton spectrum of elementary excitations very weakly depends on temperature up to the temperature $T>T_\lambda = 2.17 °K$ of the superfluid-to-normal state transition for all values of momenta, including the phonon and roton spectral regions. Furthermore, phonon--roton excitations also exist at the temperature $T>T_\lambda$ where liquid helium is in the normal state [39]. There are also experimental results on inelastic neutron scattering in liquid metals, where the phonon--roton spectrum of elementary excitations [40--42] was detected. On this basis the essential conclusion has been formulated in [45], where the phonon-roton spectrum of liquid rubidium was calculated and compared with the respective experimental results (see Fig.2 [45]). The rotons are not the specific excitations in superfluid helium, but exists as also phonons in different liquids. The continuous phonon-roton branch of excitations is a universal branch of excitations in liquid state of matter. Similar excitations were also experimentally detected in the two-dimensional Fermi liquid [46].

Let us pay attention that the Landau superfluidity criterion (1), (2) is in fact the superfluidity breakdown criterion, since it initially assumes the existence of superfluidity. Otherwise, if we take into account that well-defined acoustic elementary excitations (phonons) exist in any liquid, we would obtain a nonzero critical velocity equal to the speed of sound in a corresponding liquid in the absence of the other branch of excitations, giving a zero critical velocity. This consideration also shows the necessity of the existence of several branches of elementary excitations.

In this case, according to the Landau theory [1, 2] the helium superfluidity at temperatures $T>T_\lambda$ is caused by the presence of the superfluid component whose density, according to available experimental data, depends strongly on temperature (see, e.g., [10]). At the same time, the dependence of the critical velocity on temperature in the Landau theory is not discussed [47].

\section{Generalization of the Landau superfluidity criterion}

To clarify the situation, it should be taken into account that the spectrum of elementary excitations in the normal component is a function of not only the momentum $p$, but also the thermodynamic parameters of the system under consideration, e.g., the temperature $T$, i.e.,
\begin{eqnarray}
\varepsilon^{(\alpha)}=\varepsilon^{(\alpha)}(p,T). \label{A3}
\end{eqnarray}
This statement is a direct consequence of the general definition of elementary excitations as weakly damping poles of the analytic continuation of Green's temperature functions (see, e.g., [48]). Hence, the critical velocity $V_{cr}$ determined from relation (2) is also a function of thermodynamic parameters,$V_{cr}=V_{cr}(T)$. We further take into account that the superfluidity phenomenon is absent at the temperature $T>T_\lambda$, i.e., the liquid is normal.
Therefore, we should claim that
\begin{eqnarray}
V_{cr}(T>T_\lambda)=0. \label{A4}
\end{eqnarray}
Statement (4) is confirmed by experimental data on measurements of the critical velocity $V_{cr}^{exp}$ as the function of temperature for superfluid helium flowing through micrometer holes [49, 50].
Thus, we can formulate the generalized Landau superfluidity criterion exactly as the superfluidity criterion, rather than the superfluidity breakdown criterion, in the following form: if the spectra of elementary excitations in liquid, taking into account (2), (3), satisfy the conditions
\begin{eqnarray}
V_{cr}>0  \qquad \mbox{for}\qquad T<T_\lambda, \qquad  V_{cr}=0  \qquad \mbox{for}\qquad T>T_\lambda, \label{A5}
\end{eqnarray}
then the corresponding liquid at temperatures $T<T_\lambda$ is superfluid; therewith, the superfluidity breakdown occurs at velocities $V > V_{cr}$.

Taking into account the above-discussed experimental data, phonon--roton elementary excitations do not satisfy the superfluidity condition (5). Hence, at least one branch of elementary excitations, in addition to the phonon--roton branch, should satisfy condition (4) in superfluid helium.
To provide conditions (4), the Landau superfluidity theory [1, 2] contains an appropriate quantity, i.e., the superfluid component density $n_s$, for which the condition
\begin{eqnarray}
n_s (T>T_\lambda)=0. \label{A6}
\end{eqnarray}
is satisfied.
\begin{figure}[h]
\centering\includegraphics[width=6cm]{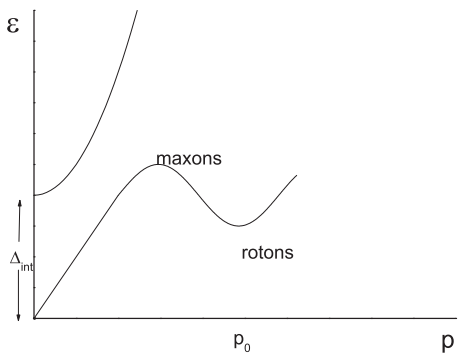}
\caption{{\protect\footnotesize {Phonon--roton spectrum and the expected spectrum of elementary excitations with the gap $\Delta_{int}$ (helons [53]) depending on the interaction and temperature at $p = 0$ . At the transition temperature, the gap vanishes, $\Delta_{int}(T=T_\lambda)=0$}}}\label{Fig.3}
\end{figure}

Then, we can assume that the energy spectrum of sought elementary excitations is defined by the quantity $n_s(T)$ at temperatures $T>T_\lambda$. For dimension reasons, we can construct several energy dimension quantities based on the superfluid component density $n_s(T)$, in particular, $\hbar^2 n_s^{2/3}/m$  and  $\hbar^2 L n_s/m$, where L is the so-called scattering length which is completely defined by the interparticle interaction potential of helium atoms.

As an appropriate candidate for the role of elementary excitations whose energy spectrum strongly depends on temperature, the so-called second sound which is a characteristic feature of superfluid helium can be considered. The speed of the second sound is directly related to the superfluid component density (see [1, 6, 10] for more details). However, the question arises, whether temperature waves (or entropy waves) of the second sound can be considered as elementary excitations similarly to phonon--roton excitations which characterize the energy spectrum of an equilibrium liquid.
\begin{figure}[h]
\centering\includegraphics[width=6cm]{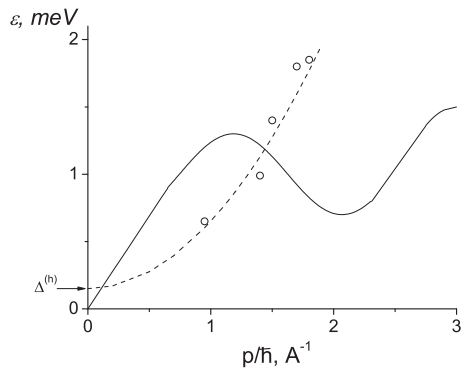}
\caption{{\protect\footnotesize {Position of the dynamic structure factor maximum of superfluid helium at 1.2 K. The solid line is the experimental phonon-roton spectrum, the circles are the positions of "quasi-free" maxima from [51,52],  the dashed curve is the assumed one for the helon excitations in HeII [53].}}}\label{Fig.4}
\end{figure}

In our opinion, temperature waves of the second sound cannot be considered as elementary excitations in the meaning adopted in the Landau theory. Here we have in mind that the energy spectrum of elementary excitations is controlled by poles of equilibrium Green's functions, which, in particular, manifest themselves in the dynamic structure factor $S(p,\omega)$.  At the same time, the second sound is a consequence of the interaction of elementary excitations in the non-equilibrium state of superfluid helium. In particular, the second sound was not detected in the experiments on inelastic neutron scattering, since its excitation requires external heating with varying temperature.

Furthermore, if the second sound is considered as one of the branches of elementary excitations, according to the Landau superfluidity criterion, the superfluidity breakdown would occur due to temperature fluctuations, which, to our knowledge, is not experimentally observed. In this case, the speed of the second sound at a temperature of $1,8 °K$ 1.8 K is $~20$ m/s [10] which also does not agree with the experimental data on the critical velocity.

The question arises about any experimental information which would confirm the existence of one more, along with the phonon--roton, branch of elementary excitations which would be consistent with the generalized Landau criterion. In our opinion, such is the branch of excitations, corresponding to the so-called "quasi-atomic" scattering can exist in the form which is shown on (the upper curve on Fig. 3). The dependence of the energy spectrum of these excitations which were detected in the experiments on inelastic neutron scattering [51, 52] is close to the spectrum of the free helium atom and is monotonic (see Fig. 4). In [53] we showed that such type excitations, which we called helons, for large transferred momenta is closed to the  "quasi-atomic" scattering, but has a gap at $p=0$ (Fig. 3). Helons can provide singularity in heat capacity of superfluid helium at the transition temperature $T_\lambda$. Numerous attempts were undertaken to construct a microscopic description of such a branch of excitations in the model of the weakly nonideal boson gas using the Bose--Einstein condensate [54--57]. However, from the viewpoint of the generalized Landau criterion, the behavior of this spectrum at small momenta is most significant due to its monotonicity. The Landau criterion validity requires that the corresponding spectrum would be characterized at small momenta either by a linear region or a gap. Unfortunately, experimental values for the spectrum of "single-atom" scattering are currently available only in the region of momenta characterized by wave vectors $k>1.5$ $A^{-1}$ [51, 52].

\section{Conclusions}

In this study, based on an analysis of experimental data, it is shown that, in addition to phonon--roton excitations in superfluid helium, at least one branch of elementary excitations whose energy spectrum strongly depends on temperature should necessarily exist. The Landau superfluidity criterion was generalized for several branches of elementary excitations, taking into account that the critical velocity should vanish during the phase transition of liquid helium from the superfluid state to the normal state. It was assumed that the validity of the generalized Landau criterion in superfluid helium is provided by the branch of "quasi-atomic" excitations, which was detected in the experiments on inelastic neutron scattering.

For this reason, it become particularly urgent to experimentally confirm the existence of elementary excitations in superfluid helium, whose energy spectrum is directly related to the superfluid component density and provides the fulfillment of the generalized Landau superfluidity criterion (5). In the experiments on inelastic neutron scattering, of most interest is the study of small momenta.

\section*{Acknowledgment}

The authors are grateful to participants of seminars of the Theoretical department of the Joint Institute for High Temperatures and the Theoretical department of the A.M. Prokhorov General Physics Institute for discussions of the results. This study was supported by the Russian Foundation for Basic Research, projects nos. 12-08-00822-a and 12-02-90433-Ukr-a. S.A. Trigger acknowledges the support of the Netherlands Organization for Scientific Research (NWO).

\end{document}